\documentclass[11pt]{article}
\usepackage[margin=1in]{geometry}
\usepackage{hyperref}
\usepackage{amsfonts}
\usepackage{fancyhdr}
\usepackage[dvipsnames]{xcolor}
\usepackage{comment}
\usepackage{graphicx} 
\usepackage{times}
\usepackage{advdate}
\usepackage{amsmath}
\usepackage{changepage}

\date{}
 
\begin{document}

\title{\LARGE
Survey on Multi-Agent Q-Learning frameworks for resource management in wireless sensor network\\
}

\author{Arvin Tashakori\\ Department of Electrical and Computer Engineering\\University of British Columbia 
}

\maketitle

\begin{abstract}
This report aims to survey multi-agent Q-Learning algorithms, analyze different game theory frameworks used,  address each framework's applications, and report challenges and future directions. The target application for this study is resource management in the wireless sensor network.

In the first section, the author provided an introduction regarding the applications of wireless sensor networks. After that, the author presented a summary of the Q-Learning algorithm, a well-known classic solution for model-free reinforcement learning problems. 

In the third section, the author extended the Q-Learning algorithm for multi-agent scenarios and discussed its challenges.

 In the fourth section, the author surveyed sets of game-theoretic frameworks that researchers used to address this problem for resource allocation and task scheduling in the wireless sensor networks. Lastly, the author mentioned some interesting open challenges in this domain.

\end{abstract}

\section{Introduction}\label{sec:intro}

In the past decades, we have seen great progress in the fields of Internet of Things (IoT) and Wireless Sensor Networks (WSN) \cite{derakhshan2019review}. There are a variety of applications that employs wireless sensor networks, including:

\begin{itemize}
\item Healthcare: which includes constantly vital signs monitoring of severely ill patients and elderly care, and may act depending on the situation.  For example, suppose a patient with diabetes type I needs constant blood sugar monitoring \cite{choi2005portable}.

\item Remote sensing, monitoring: various applications from industrial level control to environmental management, including controlling wildfires \cite{shah2013distributed}.

\item Anomaly and fault detection: This includes lots of industrial and healthcare applications—for instance, fall detection \cite{derakhshan2019review}.

\end{itemize}

Wireless sensor network provides online monitoring capabilities in situations that are not accessible (for example: controlling the temperature of a nuclear reactor, or invasive brain, or muscular signal monitoring).

Usually, Wireless sensor nodes are heterogeneous, energy-constrained, and tend to operate in dynamic and unclear situations. In these situations, nodes need to learn how to cooperate over tasks and resources (including power and bandwidth). It implies that we design a framework that allows wireless sensor nodes to adapt to the new situation. In these scenarios, Reinforcement Learning is an immeasurable solution. \cite{shah2013distributed}.

Recently, reinforcement learning (RL) becomes a trend in various autonomous decision-making tasks, whether sequential or simultaneous. For example, tasks related to solving strategic games, or sensor and communication networks, finances, social science, etc. \cite{zhang2019multi}.

Multi-agent Reinforcement Learning has shown its success in different research areas in wireless sensor networks, including resource allocation, localization, caching, data offloading \cite{althamary2019survey}.

Multi-agent Reinforcement Learning consists of several agents that interact with an environment and, based on that interaction, receive rewards. In order to model a wireless sensor network with reinforcement learning, we call wireless sensor nodes agents, the environment can be considered where they are located, or other nodes can be considered as the environment that they tend to interact with over time \cite{shah2013distributed,shah2007distributed}.

In reinforcement learning, we have an environmental states which in our case can be sets of measurement that nodes are doing, their battery level, availability of frequency spectrum \cite{shah2013distributed,cheng2017multi}. Based on how we define this set, it's size can exponentially grows \cite{shah2013distributed,shah2007distributed}.

Another metrics that need to be addressed is action list. For our purpose, nodes can receive or transmit an specific package \cite{cheng2017multi}, or even execute a certain tasks \cite{shah2013distributed}.

Lastly, we need to define how we set up the reward function. In most papers, to the best of author's knowledge, there are two types of reward function that has been explored; a) internal reward, where agents based on some of internal variables, like energy usage, define a reward function for itself; b) external reward, where agents receive a certain reward from central controller, or other nodes (for example, acknowledgement that package was successfully received) \cite{shah2013distributed}.

Multi-agent reinforcement learning problem is a broad topic. This study mainly consider solutions related to the Q-Learning, one of the classic solutions for scenarios that there is no available model for the environment. 

In order to model the environment, Q-Learning consider the environment to act as a Markov Decision Process where model the environment with sets of states, probability function based on the current state, agent's action and the next state. In the next section, the generalized definition of the Markov Decision Process, Stochastic Games, where we have multiple agents who interact with the same environment is defined.

\section{Q-Learning Background}

In order to have consistency over the report, author defined the list of main parameters in this report using notations of \cite{zhang2019multi}. An Stochastic Game is a tuple of:
\begin{itemize}
\item $\mathcal{A}=\{A^1,...,A^N\}$ is the set of agents where $N>1$.
\item $\mathcal{S}=S_0 \times S_{A^1}  \times ... \times S_{A^N}$ denotes state space which is composed of local states spaces and the shared state space.
\item $U=U_{A^1}\times...\times U_{A^N}$ is action space for each agent
\item $\mathcal{P}:\mathcal{S}\times U \times \mathcal{S}\to [0,1]$ is the transition function.
\item $\mathcal{R}=R_{A^1},...,R_{A^N}$ where $R_{A^i}:\mathcal{S}\times U \times \mathcal{S}\to \mathbb{R}$ is a real-valued reward function for agent i
\end{itemize}

In  Q-Learning,  the agent tries to find the optimum policy by iteratively interacting with the environment. In each step, the agent first observes the environment state, considering the environmental state's full observability, which is modeled using Markov Decision Process (MDP) and based on his current policy function. It decides to take any action that changes the environment state to maximize its expected accumulated reward. Based on the reward that it gets and observing the next state, it updates its belief from the environment.

In order to mathematically discuss the Q-Learning function, we first need to define action-value and value functions:

\begin{equation}
Q_\pi (s,u)= \mathbb{E} \left[ \sum_{t\geq 0} \gamma^t R(s_t,u_t,s_{t+1})| u_t\sim\pi(.|s_t),u_0=u,s_0=s\right]
\label{q}
\end{equation}

\begin{equation}
V_\pi (s)= \mathbb{E} \left[ \sum_{t\geq 0} \gamma^t R(s_t,u_t,s_{t+1})| u_t\sim\pi(.|s_t),s_0=s\right]
\label{v}
\end{equation}

The action-value function (or Q function, Equation \ref{q})  tells us the expected accumulated reward if we start from state s and take action u from the sets of available actions, where The value function (or V function, Equation \ref{v}) represents the expected accumulated rewards that you can expect if you start from state s.

The discount factor, $0 \leq \gamma \leq 1$ shows how much the agent prefers instant gratification and would like to stick with what it learned, compare to exploring new situations and taking a risk. 

If we know the optimal action-value function, then it is evident that we can calculate the optimum policy as follows:

\begin{equation}
\pi^*(u|s)=\frac{Q^*_\pi (s,u)}{\sum_{u_j\in U} Q^*(s,u_j)}
\end{equation}

In Q-Learning, the agent iteratively starts to interact with the environment. In each step, based on the state that it starts, the action that it takes, the reward that it gets, and the state that it gets, it iteratively updates its estimation of value and action-value function (Equation \ref{qlearning}). In the equation \ref{qlearning}, $\alpha$ denotes the learning rate. The goal of Q-Learning is to iteratively converge to the optimal action-value function and value function as follows (Equation \ref{qlearning}):

\begin{equation}
\begin{cases}
Q_{k+1}(s_k,u_k) \gets (1-\alpha)Q_{k}(s_k,u_k) + \alpha \left[ R(s_k,u_k,s_{k+1}) +\gamma V_k(s_{k+1})\right]\\
V_{k+1}(s_k) \gets max_{u_k \in U} Q_{k+1}(s_k,u_k)
\end{cases}
\label{qlearning}
\end{equation}

In multi-agent reinforcement learning, the action-value and value functions have a complicated definition because they are not just the function of an agent's decisions. An agent can not unilaterally define them. That is why researchers in this field employ game theory frameworks to address it.

\section{Extending Q-Learning for Multi-agent scenario}

\begin{figure}[t]

\centering
\includegraphics[width=0.7\textwidth]{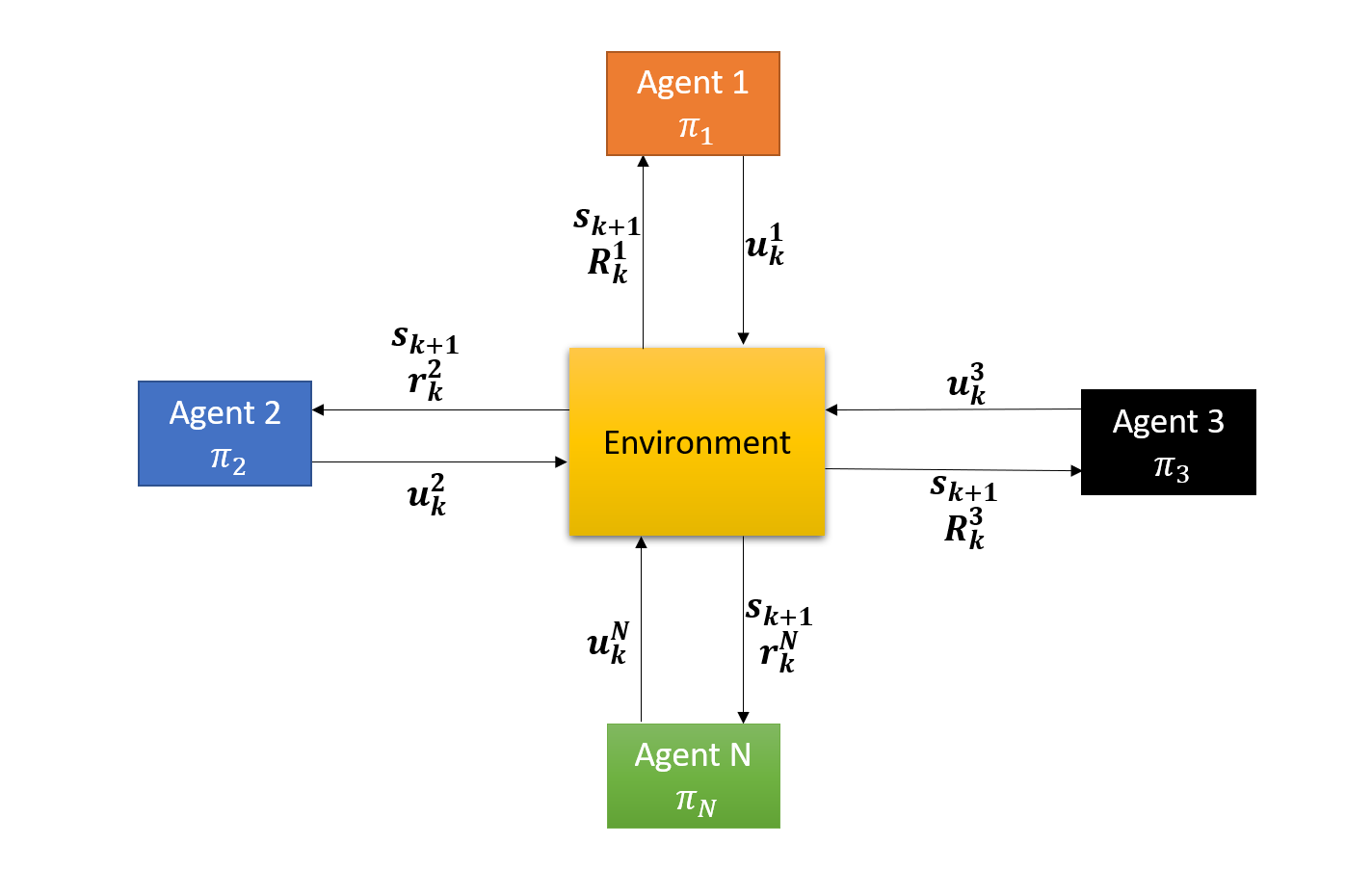}
\caption{An illustration of how agents interact with the environment in the multi-agent reinforcement learning \cite{foerster2018deep}.}
\label{fig:birds}
\end{figure}

Let's consider the case that we have multiple agents who interact with the same environment (Figure \ref{fig:birds}). The obvious solution is to consider that we have independent learners interacting with a passive environment and add index i for the action-value, value, and reward functions.
\begin{equation}
\begin{cases}
Q^i_{k+1}(s_k,u_k) \gets (1-\alpha)Q^i_{k}(s_k,u_k) + \alpha \left[ R^i(s_k,u_k,s_{k+1}) +\gamma V^i_k(s_{k+1})\right]\\
V^i_{k+1}(s_k) \gets max_{u_k \in U^i} Q^i_{k+1}(s_k,u_k)
\end{cases}
\end{equation}

There are several problems with this approach: 

First of all, In this scenario, agents can selfishly try to maximize their expected accumulated reward without considering other agents' actions. 

Secondly, it is essential to consider that we do not have a passive environment in most scenarios. In other words, agents can not unilaterally maximize their expected accumulated reward without considering other agent's actions.

Lastly, the definition of the value function is not valid anymore. Specifically, we can not update the expected accumulated reward by maximizing the action-value function over sets of available actions for agent i.

To address the first and second problem, we can add the other agents' action to the action-value and reward function (Equation \ref{eq1}).

\begin{equation}
\begin{cases}
Q^i_{k+1}(s_k,\overrightarrow{u_k}) \gets (1-\alpha)Q^i_{k}(s_k,\overrightarrow{u_k}) + \alpha \left[ R^i(s_k,\overrightarrow{u_k},s_{k+1}) +\gamma V^i_k(s_{k+1})\right]\\
V^i_{k+1}(s_k) \gets max_{u_k \in U^i} Q^i_{k+1}(s_k,\overrightarrow{u_k})
\end{cases}
\label{eq1}
\end{equation}

Now, The update rule for the action-value is acceptable. However, we still need to find how we can estimate the value function in each state. 

In the next section, author discussed approaches to find optimal value functions.

\section{Approaches to find optimal value functions}

Generally There are two main approaches to update the value function, a) using Stochastic games framework, which is a generalized-form of Markov games to multiple agents simultaneously interacting with the same environment \cite{zhang2019multi}; b) Using Extensive-form games to mostly model sequential action taking scenarios \cite{zhang2019multi}.

To the best of the author's knowledge, there is no study on using extensive form games to address multi-agent Q learning. However, based on the author's findings, we can categorized the approaches to find the optimal value function in the wireless sensor network resource management applications into four main frameworks.

\subsection{Independent Agents}

In \cite{shah2007distributed,shah2013distributed}, authors push forth the idea of independent learners as a solution for multi-agent Q-Learning algorithm in wireless sensor network resource management problem. Authors claimed that although it is more accurate to train the sensor nodes as a joint action learner, in most scenarios, the performance of agents in both frameworks are nearly the same.

They suggested that this approach will reduce the cost of training, either the whole network or simply one new sensor node, and the need for communication between the nodes.

In my opinion, there are two situations that their approach won't work:

\begin{itemize}
    \item When we need a strict coordination task between our agents for a specific task. 
    \item one of the main challenges that I've found is when there is a delay between the action that the agent takes and the reward that it gets. For instance, when the agent needs to wait for a certain acknowledgment from the receiver, our node can not connect the delayed reward to its action.
\end{itemize}

\subsection{Stochastic games}

The main and classic approach to model, the multi-agent Q-Learning problems, are using the framework of the Stochastic game. Both \cite{cheng2017multi, shoham2003multi} have suggested three algorithms as the most successful ones to update the value function: NashQ-Learning, Friend and Foe Q-Learning, and Minimax Q-Learning.

Authors in all of these three approaches have shown that under certain situations, the action-value function converges to the optimal value.

The main challenge in this framework is that due to the curse of dimensionality, it is hard to train a large number of agents \cite{oroojlooyjadid2019review}.

Based on \cite{shoham2003multi}, we can update the value function in two zero-sum agents as follows:

\begin{equation}
\begin{cases}
Q^i_{k+1}(s_k,\overrightarrow{u_k}) \gets (1-\alpha)Q^i_{k}(s_k,\overrightarrow{u_k}) + \alpha \left[ R^i(s_k,\overrightarrow{u_k},s_{k+1}) +\gamma \Phi^i_{MINMAX}(s_{k+1})\right]\\
\Phi^i_{MINMAX}(s_{k+1}) \gets max_{u_k^i} min_{u_k^{-i}} \sum_{u_k^i} P_i(u^i)Q^i_k(u_k)
\end{cases}
\end{equation}

When we have general sum scenarios, for each agent, we have sets of friends and sets of foes. based on that assumption, we can update the value function as below \cite{shoham2003multi}:

\begin{equation}
\begin{cases}
Q^i_{k+1}(s_k,\overrightarrow{u_k}) \gets (1-\alpha)Q^i_{k}(s_k,\overrightarrow{u_k}) + \alpha \left[ R^i(s_k,\overrightarrow{u_k},s_{k+1}) +\gamma \Phi^i_{FOF}(s_{k+1})\right]\\
\Phi^i_{FOF}(s_{k+1}) \gets max_{u_k^j \in friends} min_{u_k^j\in foes} \sum_{u_k^j} 	\prod_{u_k^j} \pi^m(u^m_k)Q^i_k(u_k)
\end{cases}
\end{equation}

The more general solution is called NashQ-Learning, where it updates the value function using \cite{shoham2003multi}:

\begin{equation}
\begin{cases}
Q^i_{k+1}(s_k,\overrightarrow{u_k}) \gets (1-\alpha)Q^i_{k}(s_k,\overrightarrow{u_k}) + \alpha \left[ R^i(s_k,\overrightarrow{u_k},s_{k+1}) +\gamma \Phi^i_{NASH}(s_{k+1})\right]\\
\Phi^i_{NASH}(s_{k+1}) \gets \mbox{action value at the equilibrium of the agent i for the next state}
\end{cases}
\end{equation}

\subsection{Stackelberg games}

Although most papers have employed frameworks where agents are taking actions simultaneously, authors in \cite{sengupta2020multi,cheng2017multi} used the Stackelberg games framework as a solution for certain wireless sensor network resource management problem. 

Based on my findings, this approach is more meaningful when taking simultaneous actions cost more for some agents. For example: 
\begin{itemize}
\item Cognitive radio networks, where we have two sets of users a)primary users and b)secondary users. Primary users have priority in using the spectrum compare to secondary users. In this case, it's rational for secondary users to observe the action of primary users and then take action (for example, using the spectrum).
\item In a situation where there is a jammer in the area and users wants to communicate over the same channel, it's more logical for users to first observe the jammer's action.
\end{itemize}

The authors in \cite{cheng2017multi}, have shown that the action-value function in this scenario converge faster than the NashQ-Learning algorithm. Their algorithm can be find at Equation \ref{stack}.

\begin{equation}
\begin{cases}
Q^i_{k+1}(s_k,\overrightarrow{u_k}) \gets (1-\alpha)Q^i_{k}(s_k,\overrightarrow{u_k}) + \alpha \left[ R^i(s_k,\overrightarrow{u_k},s_{k+1}) +\gamma \Phi^i_{STACK}(s_{k+1})\right]\\
\Phi^i_{STACK}(s_{k+1}) \gets \mbox{is the Stackelberg equilibrium reward the next state}
\end{cases}
\label{stack}
\end{equation}

\subsection{mean field games}
Most of the multi-agent Q-learning algorithms are consider small sets of agents interacting with each other. It's practically impossible to solve it when the number of agents are extremely large.

One of the solutions is to simplify this problem is to approximate the action-value function to local neighbors (Equation \ref{meanfield}) \cite{yang2018mean}.

\begin{equation}
Q^i(s_k,\overrightarrow{u_k}) =\frac{1}{\|N_i\|}\sum_{j \in N_i} Q^i(s_k,u^i_k,u^j_k)
\label{meanfield}
\end{equation}

The necessary assumption  for this approach is:
\begin{itemize}
\item we need to have a large number of agents.
\item agents should sparsely take action. 
\item only actions of agents in the neighborhood of every agent effect on the agent's environmental state.
\end{itemize}

\section{Conclusions}
This article discusses a survey of the multi-agent Q-learning frameworks that have been used in the wireless sensor communication resource management problems. Researchers used three main frameworks to address multi-agent Q-Learning for resource allocation in the wireless sensor networks:

\begin{enumerate}

\item They considered wireless nodes as independent learners.
\item They used the framework of the Stochastic game to model joint learner scenarios.
\item For cases where we have one leader and several followers, the authors showed that the convergence of to optimal action-value function is faster. 

\end{enumerate}

To the best of the author's knowledge, there is no work in using a mean-field approximation to model cases where we have many users, although its promising results in the literature.

\bibliographystyle{abbrv}
\bibliography{myproject_bib}
\end{document}